\documentclass[10pt]{article}
\usepackage{opex2}
\usepackage{epsfig}
\usepackage{amsmath}
\usepackage{amssymb}
\usepackage{mathrsfs}
\begin{document}

\title{Photonic crystal fiber design\\based on the $V$--parameter}

\author{Martin Dybendal Nielsen$^{1,2*}$ and Niels Asger Mortensen$^{1}$}

\address{$^1$Crystal Fibre A/S, Blokken 84, DK-3460 Birker\o d, Denmark\\
$^2$COM, Technical University of Denmark,\\DK-2800 Kongens Lyngby, Denmark}

\email{$^*$mdn@crystal-fibre.com}

\begin{abstract}
Based on a recent formulation of the $V$--parameter of a photonic crystal fiber we provide numerically based empirical expressions for this quantity only dependent on the two structural parameters --- the air hole diameter and the hole-to-hole center spacing. Based on the unique relation between the $V$--parameter and the equivalent mode field radius we identify how the parameter space for these fibers is restricted in order for the fibers to remain single mode while still having a guided mode confined to the core region.
\end{abstract}


\section{Introduction}

Theoretical descriptions of photonic crystal fibers (PCFs) have traditionally been restricted to numerical evaluation of Maxwell's equations. In the most general case, a plane wave expansion method with periodic boundary conditions is employed~\cite{johnson2001} while other methods, such as the multipole method \cite{white2002}, take advantage of the localized nature of the guided modes and to some extend the circular shape of the air-holes. The reason for the application of these methods is the relatively complex dielectric cross section of a PCF for which rotational symmetry is absent. 
 
The aim of this work is to provide a set of numerically based empirical expressions describing the basic properties such as cutoff and mode-field radius of a PCF based on the fundamental geometrical parameters only.

\begin{figure}[b!]
\begin{center}
\epsfig{file=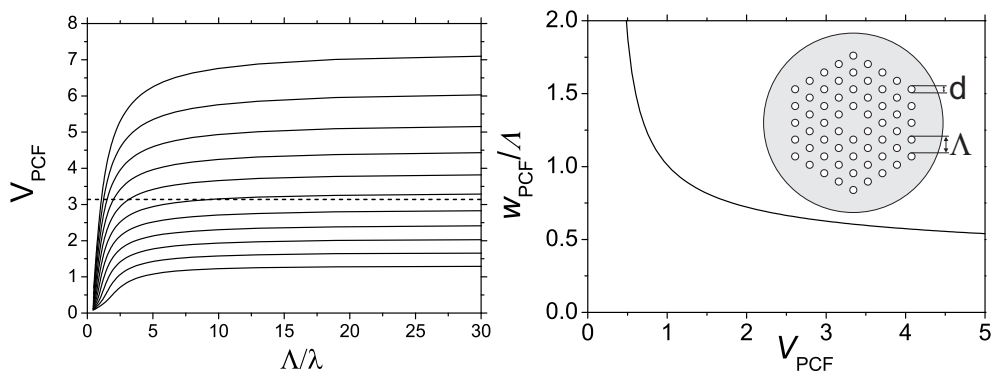, width=0.99\textwidth,clip}
\end{center}
\caption{left panel shows $V_{\rm PCF}$ calculated from Eq.~(\ref{fig1}) for $d/\Lambda$ ranging from 0.20 (lowest curve) to 0.70 in steps of 0.05. The dashed line indicates $V_{\rm PCF} =\pi$. The right panel shows the relative equivalent mode-field radius, $w_{\rm PCF}/\Lambda$ plotted as function of $V_{\rm PCF}$ for each of the 9 curves in the left panel. The inset shows a schematic drawing of the considered PCF structure.} 
\label{fig1}
\end{figure}

\section{Fiber geometry and numerical method}

We consider the fiber structure first studied by Knight {\it et al.} \cite{knight1996} and restrict our study to fibers that consist of pure silica with a refractive index of 1.444. The air holes of diameter $d$ are arranged on a triangular grid with a pitch, $\Lambda$. In the center an air hole is omitted creating a central high index defect serving as the fiber core. A schematic drawing of such a structure is shown in the inset of the right panel in Fig.~\ref{fig1}.

Depending on the dimensions, the structure comprises both single- and multi-mode fibers with large mode area as well as nonlinear fibers. The results presented here cover relative air hole sizes, $d/\Lambda$, from 0.2 to 0.9 and normalized wavelengths, $\lambda/\Lambda$, from around 0.05 to 2. The modeling is based on the plane-wave expansion method with periodic boundary conditions~\cite{johnson2001}. For the calculations of guided modes presented the size of the super cell was $8\times 8$ resolved by $256\times 256$ plane waves while for calculations on the cladding structure only, the super cell was reduced to a simple cell resolved by $32\times 32$ planes waves.

\begin{figure}[t!]
\begin{center}
\epsfig{file=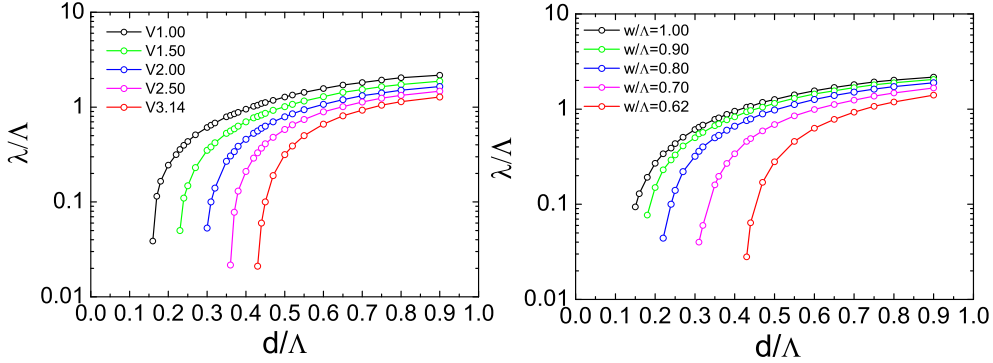, width=0.99\textwidth,clip}
\end{center}
\caption{The left panel shows curves for constant values of $V_{\rm PCF}$ in a normalized wavelength versus relative hole-size plot. The open circles indicate calculated data points with full lines to guide the eye. Similarly, the right panel shows curves for constant relative equivalent mode-field radius.} 
\label{fig2}
\end{figure}

\section{The $V$--parameter and the relative mode-field radius}

When attempting to establish a simple formalism for the PCF it is natural to strive for a result similar to the $V$--parameter known from standard fibers \cite{snyder,marcuse1978}. However, a simple translation is not straight forward since no wavelength-independent core- or cladding index can be defined. Recently, we instead proposed a formulation of the $V$--parameter for a PCF given by~\cite{mortensen2003c} 

\begin{equation}\label{VPCF}
V_{\rm PCF}=2\pi \frac{\Lambda}{\lambda}\sqrt{n_{\rm FM}^2(\lambda)-n_{\rm FSM}^2(\lambda)}
\end{equation} 
Although this expression has the same overall mathematical form as known from standard fibers, the unique nature of the PCF is taken into account. In Eq.~(\ref{VPCF}), $n_{\rm FM}(\lambda)$ is the wavelength dependent effective index of the fundamental mode (FM) and  $n_{\rm FSM}(\lambda)$ is the corresponding effective index of the first cladding mode in the infinite periodic cladding structure often denoted the fundamental space filling mode (FSM). For a more detailed discussion of this expression and its relation to previous work we refer to Ref.~\cite{mortensen2003c} and references therein. We have recently argued that the higher-order mode cut-off can be associated with a value of $V_{\rm PCF} =\pi$  \cite{mortensen2003c} and showed that this criterion is indeed identical to the single-mode boundary calculated from the multipole method \cite{kuhlmey2002}. Recently the cut off results have also been confirmed experimentally \cite{folkenberg2003}. Further supporting the definition of $V_{\rm PCF}$ is the recent observation \cite{nielsen2003b} that the relative equivalent mode field radius of the fundamental mode, $w_{\rm PCF}/\Lambda$ as function of $V_{\rm PCF}$ fold over a single curve independent of $d/\Lambda$. The mode field radius $w_{\rm PCF}$ is defined as $A_{\rm eff}=\pi w_{\rm PCF}^2$ and corresponds to the $1/e^2$ width of a Gaussian intensity distribution with the same effective area, $A_{\rm eff}$, as the fundamental mode itself \cite{nielsen2003b}.

In the left panel of Fig.~\ref{fig1}, calculated curves of $V_{\rm PCF}$ as function of $\Lambda/\lambda$ are shown for $d/\Lambda$ ranging from 0.20 to 0.70 in steps of 0.05. In general, all curves are seen to approach constant levels dependent on $d/\Lambda$. The horizontal dashed line indicates the single-mode boundary $V_{\rm PCF} =\pi$. In the right panel, $w_{\rm PCF}/\Lambda$ is plotted as function of $V_{\rm PCF}$ for each of the 9 curves in the left panel and as seen all curves fold over a single curve. An empirical expression for $w_{\rm PCF}/\Lambda$ can be found in Ref.~\cite{nielsen2003b}. The mode is seen to expand rapidly for small values of $V_{\rm PCF}$ and the mode-field radius saturates toward a constant value when $V_{\rm PCF}$ becomes large. In fact, it turns out that $w_{\rm PCF}/\Lambda\simeq 1.00$ for $V_{\rm PCF} =1$ and $w_{\rm PCF}/\Lambda\simeq 0.62$ for $V_{\rm PCF} =\pi$. In the left panel of Fig.~\ref{fig2}, curves corresponding to constant values of $V_{\rm PCF}$ are shown in a $\lambda/\Lambda$ versus $d/\Lambda$ plot. In the right panel, curves of constant $w_{\rm PCF}/\Lambda$ is shown, also in a $\lambda/\Lambda$ versus $d/\Lambda$ plot. Since there is a unique relation between $w_{\rm PCF}/\Lambda$ and $V_{\rm PCF}$ \cite{nielsen2003b} the curves naturally have the same shape.

\begin{figure}[t!]
\begin{center}
\epsfig{file=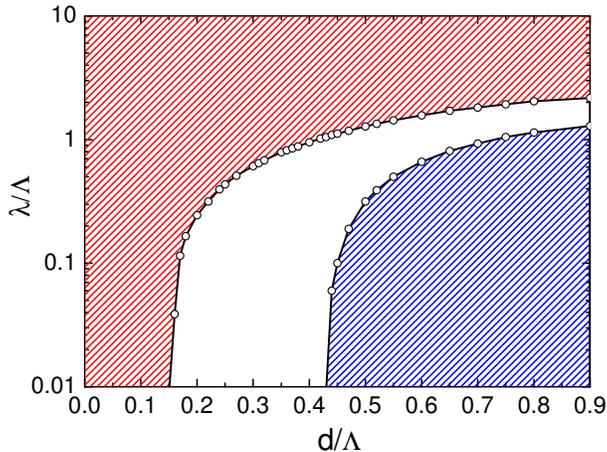, width=0.6\textwidth,clip}
\end{center}
\caption{Plot of the parameter space in terms of relative hole size and normalized wavelength divided into three regions by the boundaries defined by $V_{\rm PCF} = 1$ and $V_{\rm PCF} =\pi$. In the upper red area the mode penetrates deeply into the cladding region and in lower blue region the structure supports a higher-order mode.} 
\label{fig3}
\end{figure}

When designing a PCF any combination of $d$ and $\Lambda$ is in principle possible. However, in some cases the guiding will be weak causing the mode to expand beyond the core and into the cladding region~\cite{white2001,kuhlmey2002c} corresponding to a low value of $V_{\rm PCF}$. In the other extreme, the confinement will be too strong allowing for the guiding of higher-order modes \cite{mortensen2003c,kuhlmey2002}. Since both situations are governed by $V_{\rm PCF}$ the design relevant region in a $\lambda/\Lambda$ versus $d/\Lambda$ plot can be defined. This is done in Fig.~\ref{fig3} where the low limit is chosen to be $V_{\rm PCF} = 1$ where $w_{\rm PCF}/\Lambda\simeq 1$. How large a mode that can be tolerated is of course not unambiguous. However, for $w_{\rm PCF}\sim \Lambda$ leakage-loss typically becomes a potential problem in PCFs with a finite cladding structure. In non-linear PCFs it is for dispersion reasons often advantageous operating the PCF at $V_{\rm PCF} \lesssim 1$ and then a high number of air-hole rings is needed to achieve an acceptable level of leakage loss \cite{reeves2002}. 

Finally, we note that the practical operational regime is also limited from the low wavelength side. In Ref.~\cite{mortensen2003b} a low-loss criterion was formulated in terms of the coupling length $z_c=\lambda/[n_{\rm FM}(\lambda)-n_{\rm FSM}(\lambda)]$ between the FM and the FSM. In general scattering-loss due to longitudinal non-uniformities increases when $z_c$ increases and a PCF with a low $z_c$ will in general be more stable compared to one with a larger $z_c$. Using $n_{\rm FM}+n_{\rm FSM}\approx 2n_{\rm FM} \approx 2 n_{\rm silica}$ we can rewrite Eq.~(\ref{VPCF}) as

\begin{equation}
V_{\rm PCF}\propto \frac{\Lambda}{\lambda} \sqrt{\frac{\lambda}{z_c(\lambda)}}
\end{equation}
from which it is seen that a high value of the $V$--parameter is preferred over a smaller value. In Fig.~(\ref{fig3}) it is thus preferable to stay close to the single-mode boundary ($V_{\rm PCF}\sim\pi$) but in general there is a practical lower limit to the value of $\lambda/\Lambda$ which can be realized because when $\lambda/ \Lambda \lesssim 0.1$ one generally has that $z_c\gg\lambda$ \cite{mortensen2003b}.

\begin{figure}[t!]
\begin{center}
\epsfig{file=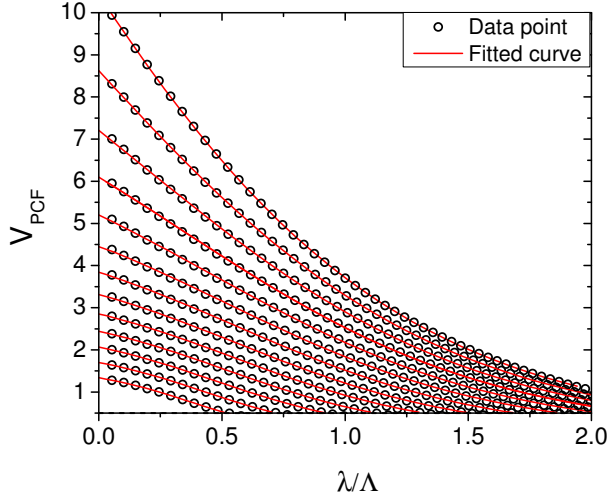, width=0.6\textwidth,clip}
\end{center}
\caption{Plot of $V_{\rm PCF}$ as a function of relative wavelength $\lambda/\Lambda$ for $d/\Lambda$ ranging from 0.20 (lowest curve) to 0.80 in steps of 0.05.} 
\label{fig4}
\end{figure}

\section{$V$--parameter expression}

Although the $V$--parameter offers a simple way to design a PCF, a limiting factor for using Eq.~(\ref{VPCF}) is that a numerical method is still required for obtaining the effective indices. In analogy with expressions for standard fibers \cite{marcuse1978} it would therefore be convenient to have an alternative expression only dependent on the wavelength, $\lambda$, and the structural parameters $d$ and $\Lambda$. In Fig.~\ref{fig4}, we show $V_{\rm PCF}$ as function of $\lambda/\Lambda$ (data are shown by open circles) for $d/\Lambda$ ranging from 0.20 to 0.80 in steps of 0.05. Each data set in Fig.~\ref{fig4} is fitted to a function of the form
\begin{subequations}\label{VPCF_fit}
\begin{equation}\label{VPCF_fit_V}
V_{\rm PCF}\left(\tfrac{\lambda}{\Lambda},\tfrac{d}{\Lambda}\right) = \frac{{\mathscr A}\left(\tfrac{d}{\Lambda}\right)}{{\mathscr B}\left(\tfrac{d}{\Lambda}\right) \times\exp\Big[{\mathscr C} \left(\tfrac{d}{\Lambda}\right)\times \tfrac{\lambda}{\Lambda}\Big]+1}
\end{equation} 
and the result is indicated by the full red lines. Eq.~(\ref{VPCF_fit_V}) is not based on considerations of the physics of the V-parameter but merely obtained by trial and error in order to obtain the best representation of calculated data with the lowest possible number of free parameters. Prior to the fit, the data sets are truncated at $V_{\rm PCF} = 0.5$ since $w_{\rm PCF} \gtrsim 2\Lambda$ in this region (see left panel in Fig.~\ref{fig1}) and the data is thus not practically relevant. In Eq.~(\ref{VPCF_fit_V}) the fitting parameters $\mathscr A$, $\mathscr B$, and $\mathscr C$ depend on $d/\Lambda$ only. In order to extract this dependency, suitable functions (again obtained by trial and error) are fitted to the data sets for $\mathscr A$, $\mathscr B$, and $\mathscr C$. We find that the data are well described by the following expressions

\begin{equation}
{\mathscr A}\left(\tfrac{d}{\Lambda}\right) = \tfrac{d}{\Lambda} + 0.457 + \frac{3.405\times \tfrac{d}{\Lambda} }{0.904- \tfrac{d}{\Lambda}}
\end{equation}

\begin{equation}
{\mathscr B}\left(\tfrac{d}{\Lambda}\right) = 0.200 \times \tfrac{d}{\Lambda} + 0.100  + 0.027\times \left( 1.045 - \tfrac{d}{\Lambda}\right)^{-2.8}
\end{equation}

\begin{equation}\label{VPCF_fit_c}
{\mathscr C}\left(\tfrac{d}{\Lambda}\right) = 0.630 \times \exp \left( \frac{0.755}{0.171 + \tfrac{d}{\Lambda}}\right)
\end{equation}
\end{subequations}
The above set of expressions, Eqs.~(\ref{VPCF_fit}), constitute our empirical expression for the $V$--parameter in a PCF with $\lambda/\Lambda$ and $d/\Lambda$ being the only input parameters. For $\lambda/\Lambda< 2$ and $V_{\rm PCF}> 0.5$ the expression gives values of $V_{\rm PCF}$ which deviates less than $3\%$ from the correct values obtained from Eq.~(\ref{VPCF}).

\section{Endlessly single-mode criterion}
 
\begin{figure}[t!]
\begin{center}
\epsfig{file=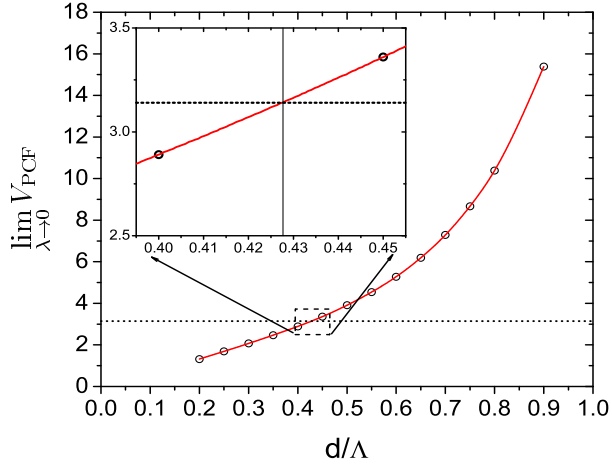, width=0.6\textwidth,clip}
\end{center}
\caption{Plot of $V_{\rm PCF}$ in the $\lambda\rightarrow 0$ limit as function of the relative air hole size (open circles). The full red line represents a fit to the data points and the horizontal dash line indicated the ESM limit $V_{\rm PCF} =\pi$. The insert shows a close-up of the intersection with the vertical line indicating the air hole size $d/\Lambda\simeq 0.43$.} 
\label{fig5}
\end{figure}

The term endlessly single-mode (ESM) refers to PCFs which regardless of wavelength only support the two degenerate polarization states of the fundamental mode \cite{birks1997}. In the framework of the $V$--parameter this corresponds to structures for which $V_{\rm PCF} < \pi$  for any $\lambda/\Lambda$ \cite{mortensen2003c}. As seen in the left panel of Fig.~\ref{fig1} this corresponds to sufficiently small air holes. However, from the plot in Fig.~\ref{fig1} it is quite difficult to determine the exact $d/\Lambda$ value for which $V_{\rm PCF} =\pi$ for $\lambda$ approaching 0. From Eq.~(\ref{VPCF_fit}) it is easily seen that the value may be obtained from

\begin{equation}\label{d_ESM}
\lim_{\lambda \rightarrow 0}V_{\rm PCF}\left(\tfrac{\lambda}{\Lambda},\tfrac{d}{\Lambda}\right) = \frac{{\mathscr A}\left(\tfrac{d}{\Lambda}\right)}{{\mathscr B}\left(\tfrac{d}{\Lambda}\right) +1} =\pi.
\end{equation}
Fig.~\ref{fig5} illustrates this equation graphically where we have extrapolated the data in Fig.~\ref{fig4} to $\lambda=0$. From the intersection of the full line with the dashed line we find that $d/\Lambda\simeq 0.43$ bounds the ESM regime. Solving Eq.~(\ref{d_ESM}) we get $d/\Lambda\simeq 0.44$ and the deviation from the numerically obtained value is within the accuracy of the empirical expression.

\section{Conclusion}

There are several issues to consider when designing a PCF. In this work we have addressed the single/multi-mode issue as well as those related to mode-field radius/field-confinement, and mode-spacing. We have shown how these properties can be quantified via the $V$--parameter. Based on extensive numerics we have established an empirical expression which facilitate an easy evaluation of the $V$-parameter with the normalized wavelength and hole-size as the only input parameters. We believe that this expression provides a major step away from the need of heavy numerical computations in design of solid core PCFs with triangular air-hole cladding.

\section*{Acknowledgments}

We thank J.~R. Folkenberg for stimulating discussion and M.~D. Nielsen acknowledges financial support by the Danish Academy of Technical Sciences.

\end{document}